# Low-loss phononic integrated circuits based on a silicon nitride-lithium niobate platform

Jun Ji[1,†,*], Joseph G. Thomas[1,†], Zichen Xi[1,†], Ruxuan Liu[2], Kinson Fang[2], Yuan Qin[1], Andreas Beling[2], Xu Yi[2], Yizheng Zhu[1], and Linbo Shao[1,3,*]

*[1] Bradley Department of Electric and Computer Engineering, Virginia Tech, Blacksburg, VA USA*
*[2]Department of Electrical and Computer Engineering, University of Virginia, Charlottesville, VA, USA*
*[3]Department of Physics and Center for Quantum Information Science and Engineering (VTQ), Virginia Tech, Blacksburg, VA, USA*

[†]J.J., J.G.T., and Z.X. contributed equally to this work.
[*]*Email: junji@vt.edu; shaolb@vt.edu*

## Abstract

Microwave-frequency acoustic waves in solids have emerged as a versatile platform for classical and quantum applications. While phononic integrated devices and circuits are being developed on various material platforms, an ideal phononic integrated circuit (PnIC) platform should simultaneously support low-loss waveguides, high-quality-factor resonators, high-performance modulators, and efficient electromechanical transducers. Here, we establish a low-loss gigahertz-frequency PnIC platform based on patterned thin-film silicon nitride on lithium niobate substrate. We develop low-loss PnIC building blocks, including acoustic waveguides with a propagation frequency-quality factor ($fQ$) product up to $4.22 \times 10^{13}$, directional couplers with controllable power transfer at an insertion loss of 3.5 dB, and ring resonators with a loaded quality factor up to 17,925. As an application, we demonstrate a 1-GHz ring-resonator-based phononic oscillator, reaching a low phase noise of -159.0 dBc/Hz at a 100-kHz offset frequency. Our low-loss PnICs could provide opportunities for microwave acoustics, quantum phononics, and integrated hybrid systems combining phonons, photons, superconducting qubits, and solid-state defects.

## Introduction

Acoustic waves in solids at microwave frequencies have attracted broad attention for both classical [1-3] and quantum [4,5] applications. Their unique properties of small footprints, long lifetime, and reduced crosstalk are widely exploited in signal processing [6,7], sensing [8,9], computing [10,11], and communication [12,13], making them a complementary platform to electronics and photonics. Meanwhile, gigahertz-frequency acoustic waves are envisioned as universal quantum transducers owing to their efficient coupling to versatile quantum systems, including superconducting qubits [14,15], optical devices [16,17], and defect centers [18,19]. They also show strong potential as more coherent and hardware-efficient quantum memories [20,21] and processors [22] for future quantum networks.

Microwave-frequency acoustic devices have been primarily used as discrete components, such as film bulk acoustic resonators (FBARs) [23-25], delay lines [26,27], and phononic-crystal nanobeams [28,29]. With the surging utility of gigahertz-frequency acoustic waves, scaling from discrete devices to large-scale integrated circuits would unlock expanded functionality and superior performance. Motivated by photonic integrated circuits [30-32] that enable large-scale optical computing and sensing systems on chip, phononic integrated circuits (PnICs) [33] are expected to offer similar capabilities by integrating key building blocks, such as interdigital transducers (IDT), phononic waveguides [34-38], modulators [39-41], directional couplers [42,43], and ring resonators [44-46], on a single piezoelectric material platform such as gallium nitride (GaN) [43-45,47-49], lithium niobate (LN) [46], and aluminum scandium nitride (AlScN) [42]. However, their applications in microwave engineering and quantum science are hindered by one or more technical challenges including high insertion loss, absence of high frequency-$Q$-product (high $fQ$) resonators, or inefficient transducers. We believe these challenges could be fully addressed by systematic engineering of each building block guided by a comprehensive understanding of acoustic-wave device dynamics and loss mechanisms. For example, our in-house optical vibrometer [50] can measure displacement profile in a high-throughput manner.

In this Article, we establish a low-loss gigahertz-frequency PnIC platform based on thin-film silicon nitride (SiN) on LN. By patterning the thin-film SiN layer, we demonstrate low-insertion-loss PnIC building blocks at 1 GHz including waveguides, directional couplers, and high-$Q$ ring resonators. Our acoustic multimode (single-mode) waveguide features a propagation loss of 3.5 dB/cm (1.9 dB/cm). Our acoustic taper from wide IDT to waveguide shows a tapering loss of < 1 dB. Our directional coupler consists of two closely spaced waveguides forming evanescent coupling. By adjusting coupling length, we can design power coupling rates from a 50/50 power splitter to a near-complete power transfer with an extinction ratio of 25 dB. The insertion loss associated with our directional coupler is 3.5 dB. Benefiting from our low-loss waveguide and coupler, our ring-like resonator supports a series of high-$Q$ modes with a free spectral range (FSR) of 0.403 MHz. Leveraging the 1,001.15 MHz acoustic-wave mode with a high $Q$ factor of 17,925 and an insertion loss of 28.2 dB, our acoustic-wave oscillator achieves a low phase noise of -159.0 dBc/Hz at a 100-kHz offset frequency, surpassing commercial signal generators. We can continuously tune the oscillation frequency from 1,000 MHz to 1,007 MHz by a combination of phase shifter control and thermal tuning.

## Results

LN has been an emerging platform for discrete phononic devices for their recently developed device fabrication [32] and promising device performances [51]. Scaling discrete devices to integrated circuits, **Fig. 1** illustrates the schematic of our PnICs on a bulk LN substrate with an etched SiN thin film on top. Surface acoustic waves (SAWs) are excited by IDTs, then routed by low-loss acoustic waveguides to interact with low-loss directional couplers, high-$Q$ resonators (e.g., ring resonators and phononic-crystal resonators [52]), Mach-Zehnder interferometers, and eventually detected by IDTs. X cut of LN is chosen since it supports both efficient SAW generation via a large electro-mechanical coupling efficiency $k^2$ (which is 4.2% at 1 GHz, details in **Supplementary Note 1**) and strong electro-acoustic modulations [39]. Alternative cuts, such as 128°Y, may be preferable for applications where maximizing $k^2$ is the primary objective.

### Low-loss phononic waveguides

Low-loss phononic waveguides are key elements of PnICs to route phonons between localized phononic building blocks such as phononic resonators. Our low-loss acoustic waveguides (**Fig. 2a**) are defined by opening a slot of width $W$ in the 300-nm-thick SiN thin film on top of bulk LN [39], with acoustic Rayleigh modes being mostly confined in LN (**Supplementary Fig. 1**).

Efficient tapering with suppressed scattering loss from wide IDTs to narrow waveguides has long been a challenge in PnICs [34,37,42,53,54]. Here, we show that both the orientation of waveguides and the tapering angle $\theta$ (**Fig. 2b**) play critical roles in mitigating the scattering loss in the taper region. On the one hand, our acoustic waveguides are oriented along a 30° angle with respect to the crystal $Z$ axis, a direction that exhibits the lowest acoustic phase velocity on the $X$-cut surface [39] and therefore provides enhanced acoustic confinement and more efficient tapering. On the other hand, a linear taper with a tapering angle $\theta \approx 4.3°$ enables efficient transition of Rayleigh wave from the wide IDT to the narrow waveguide and vice versa, as evidenced by the measured out-of-plane displacement profile (**Fig. 2c**) obtained using our in-house microwave-frequency optical vibrometer [50]. Under these conditions, the tapering loss is minimal (< 1 dB) at 1 GHz by comparing $S_{21}$ of the tapered waveguide with that of a reference IDT pair (**Fig. 2d**). We note that the taper angle was not obtained from a closed-form analytical calculation. Instead, it was chosen based on the general adiabatic transition principle that a smaller taper angle, corresponding to a longer transition length, reduces mode mismatch and scattering. In contrast, a larger tapering angle of $\theta \approx 9.4°$ showed a significant tapering loss of approximately 13 dB in electrical transmission measurements and pronounced scattering revealed by optical vibrometry (**Supplementary Fig. 2**).

We characterize the propagation loss of acoustic waveguides by measuring $S_{21}$ of straight waveguides with lengths $L$ ranging from 100 μm to 5100 μm. For large waveguide widths such as $W$ = 10 μm (**Fig. 2e**) and $W$ = 14 μm (**Supplementary Fig. 3**), high-order modes are supported (**Supplementary Fig. 1**), and the propagation loss of these multimode waveguides is around 3.5 dB/cm. For small waveguide widths such as $W$ = 6 μm, only the fundamental mode is supported (**Supplementary Fig. 1**), and the propagation loss of the single-mode waveguide is as low as 1.9 dB/cm (**Supplementary Fig. 3**). The lower propagation loss from the single-mode waveguide is attributed to two factors. On the one hand, the fundamental mode profile is mostly confined in the center of LN, leading to less scattering loss from the interaction between the modes and the rough sidewalls of SiN. On the other hand, the single-mode waveguide prohibits surface-roughness-induced or defect-induced mode conversions from the fundamental even mode to high-order odd modes that cannot be detected by the IDT. Our single-mode acoustic waveguide achieves a propagation $Q$ of 42,241 and a propagation $f \cdot Q$ product of $4.22 \times 10^{13}$, comparable to the highest value surveyed in **Table 1** ($4.40 \times 10^{13}$ Hz for an acoustic delay line). The propagation $Q$, which provides a more direct comparison across different operating frequencies and acoustic wavelengths, is defined as $(\pi f \alpha)/(\mathrm{PL} \cdot v_g)$ [55], where $f$ is the center frequency in Hz, PL is the propagation loss in dB/m, α is a constant of 20/ln(10) to convert the amplitude between Neper and decibel units, and $v_g$ is the group velocity in m/s. We note that using the $f \cdot Q$ product as a figure of merit assumes that the dominant loss is the phonon-phonon dissipation in the Akhiezer limit [56,57].

We extract the propagation loss of bent acoustic waveguides by subtracting the measured $S_{21}$ of one U-shape bend (**Fig. 2f**) and two cascaded U-shape bends (**Fig. 2h**). For a fixed waveguide width of 10 μm, the propagation loss of bent acoustic waveguides is reduced significantly from 176.5 dB/cm to 6.5 dB/cm and approaches that of the straight waveguide, as the bending radius increases from 300 μm to 1300 μm (**Fig. 2k**). At a small bending radius such as $R$ = 300 μm (**Fig. 2g**), significant acoustic wave leakage is observed from the out-of-plane displacement profiles. Acoustic power in fundamental mode is significantly coupled to the 2nd-order mode in the bent section and then inefficiently coupled back to the fundamental mode at the very end of the bend. At a large bending radius such as $R$ = 1300 μm (**Figs. 2i and 2j**), the coupling to the 2nd-order mode is suppressed throughout the propagation in the bending section (so-called adiabatic bend [58] in the integrated photonic waveguides) and a better confinement is observed. For a fixed waveguide radius of 700 μm, we observed an optimal waveguide width of 14 μm with a minimum propagation loss (**Supplementary Fig. 4**). This is likely a matched bend case [58,59] which is well-known in integrated photonic waveguides. In this case (**Supplementary Fig. 4c**), the phase of the fundamental and 2nd-order modes return in phase after the propagation in the bent section, and thus acoustic power is efficiently coupled back to the fundamental mode at the very end of the bend [59].

## Low-loss phononic directional couplers

Phononic directional couplers are essential devices for splitting and combining acoustic waves in phononic systems. Based on our low-loss acoustic waveguides, we demonstrate a low-loss phononic directional coupler with controllable power transfer (**Fig. 3a**). Our directional coupler consists of two evanescently coupled waveguides, with a coupling gap $G_c$ and a coupling length $L_c$ (**Fig. 3b**). With a fixed $G_c$ of 2 μm and adjusting the coupling length, we demonstrate controlled power transfer (**Fig. 3c**) from the input port (P1) to the through port (P2) and drop port (P3). At a coupling length of 100 μm, we obtain a 50/50 power splitter with an insertion loss of 6.5 dB from the coupler and 10 dB loss from IDT pair at each output port. In this case, the measured out-of-plane displacement profile at the through port and drop port have a similar magnitude (**Fig. 3d**). At a coupling length of 300 μm, we have a near-complete power transfer to the drop port with an insertion loss of 3.5 dB associated with the coupler and an extinction ratio of 25 dB between two output ports, which is also confirmed by measured out-of-plane displacement profile (**Fig. 3e**).

### High-*Q* phononic ring resonators

High-*Q* phononic resonators are essential for PnIC in applications such as signal generation [60,61], signal processing [62,63], and sensing [8,44]. Our resonator consists of a ring-like structure evanescently coupled to two waveguides (**Fig. 4a**). The widths of waveguides and the ring are chosen as 14 μm to minimize propagation loss. Benefiting from low tapering loss and low propagation loss of our waveguide, the ring resonator supports a series of high-*Q* acoustic modes. For a ring radius of $R$ = 1300 μm, these modes are spectrally spaced at a FSR of 0.403 MHz (**Fig. 4b**) at a group velocity $v_g$ of 3,400 m/s. The mode at $f_0$ = 1001.15 MHz simultaneously features a high loaded $Q$ factor of 17,925 and an insertion loss (from cable to cable) of -28.2 dB (**Fig. 4c**) between the input port (P1) and the drop port (P3). We extract the intrinsic $Q$ factor $Q_{in}$ and the coupling $Q$ factor $Q_c$ (per coupler) using temporal coupled mode theory (details in **Supplementary Note 2**). With a coupling gap $G_c$ = 2 μm and a coupling length $L_c$ = 100 μm, our ring resonator operates in the under-coupled regime with $Q_{in}$ = 22,393 and $Q_c$ = 179,680. The 28.2 dB insertion loss is mainly contributed by two parts: approximately 14.2 dB from the IDT pair and the access waveguides between IDTs and the waveguide-ring coupling, and approximately 14.0 dB from the under-coupled ring response. The corresponding propagation loss rate per unit length, inferred from $Q_{in}$, is estimated as $\mathrm{PL} = \pi f_0\alpha/(v_g Q_{in})$ = 3.6 dB/cm, which is consistent with the propagation loss extracted from the waveguide measurements (**Supplementary Fig. 3b**). Our ring resonator achieves an intrinsic $fQ$ product of $2.24 \times 10^{13}$, comparable to $3.4 \times 10^{13}$ of the state-of-the-art ring resonator [45].

### Low-phase-noise phononic oscillators

Similar to photonic integrated circuits [64], large-scale PnICs will require a large number of microwave-frequency signal sources that can be integrated and interconnected on a single chip. Although highly optimized discrete acoustic oscillators have achieved excellent performance [65-67], they are not well suited to densely interconnect and scale with phononic building blocks. A promising solution is to develop on-chip low-phase-noise phononic oscillators that can be interconnected through phononic waveguides. Here, we demonstrate an important step toward this goal using a hybrid phononic oscillator based on an on-chip high-*Q* acoustic ring resonator, with the gain, phase control, and output coupling implemented off chip. The ring resonator serves as the frequency-selective acoustic element of the oscillator, while the PnIC platform provides a pathway toward densely integrating acoustic oscillators with other phononic components on a single chip.

For phononic oscillators, loaded $Q$ and insertion loss of the resonator govern the phase noise performance at different spectrum regions [68]. As a proof-of-concept, the under-coupled ring resonator is chosen to maintain a loaded $Q$ close to the intrinsic $Q$, at the expense of a relatively large add–drop insertion loss. The high loaded $Q$ benefits the close-to-carrier (particularly at 1-kHz and 10-kHz) phase-noise performance by increasing the acoustic energy storage time and suppressing loop phase fluctuation [68], while the large add-drop insertion loss limits the far-offset noise floor by requiring higher loop gain.

We implemented a positive feedback loop (**Fig. 4d**, details in **Methods - Device characterizations**) between the input port and the drop port of the resonator, in which the losses are fully compensated by the gain of a low-noise amplifier (LNA). The loop phase is tuned by a phase shifter such that, at the 1001.15 MHz mode with the lowest insertion loss, the total phase delay is an integer multiple of $2\pi$ and self-oscillation is sustained. The experimental procedure used to select the desired mode and avoid mode hopping is discussed in **Supplementary Note 3**. The signal of our oscillator at a level of 5.6 dBm is coupled out from the coupler, and its phase noise is directly measured using a phase noise analyzer under ambient laboratory conditions (**Fig. 4e**). During the phase-noise measurements, neither active feedback nor dedicated passive temperature stabilization is applied. During the phase-noise measurements, the sample stage temperature is left uncontrolled under ambient laboratory conditions. For comparison, the phase noise spectra of two commercial signal generators (Rigol DSG836A and Keysight N5183B equipped with the low-phase-noise option) at 1 GHz carrier frequency are provided for comparison.

At a 1-kHz (10-kHz) offset frequency, the phase noise of our oscillator reaches -106.3 dBc/Hz (-138.4 dBc/Hz), representing an improvement of 9.5 dB (22.3 dB) over the Rigol signal generator. At these low offset frequencies, the measured phase noise exhibits a -30 dB/decade slope, which is mainly attributed to the temperature fluctuation $\Delta T$ [8]. Compared with oscillators based on phononic-crystal resonators [60] and acoustic delay lines [69], our ring resonator possesses a larger acoustic mode volume $V$, leading to a larger heat capacity $C = C_p \times \rho \times V$, and consequently reduced temperature fluctuation $\Delta T = \sqrt{k_B T^2 / C}$. Here, $C_p$ is the heat capacity at constant pressure of LN, $\rho$ is the density of LN, $k_B$ is the Boltzmann constant, and $T$ is the environmental temperature.

At larger offset frequencies, our oscillator achieves a low noise floor of -159.0 dBc/Hz at a 100-kHz offset frequency, which is 13 dB lower than that of the Keysight signal generator yet 10 dB away from that of the best reported acoustic oscillator [68]. In this regime, the phase noise is dominated by the background thermal noise at the LNA input. Reducing the resonator insertion loss would allow operation with lower amplifier gain to reach oscillation threshold, thereby reducing amplified thermal noise and further lowering the noise floor.

To benchmark our oscillators against representative radio-frequency acoustic oscillators, we use a commonly used figure of merit (FOM) that accounts for the center frequency ($f_0$) and power consumption ($P_{\mathrm{DC}}$) [70],

$$\mathrm{FOM}(\Delta f) = -\mathcal{L}_{\mathrm{dB}}(\Delta f) + 20\log\left(\frac{f_0}{\Delta f}\right) - 10\log\left(\frac{P_{\mathrm{DC}}}{1\mathrm{mW}}\right), \quad (1)$$

where $-\mathcal{L}_{\mathrm{dB}}(\Delta f)$ is phase noise expressed in decibels, and $\Delta f$ is the offset frequency. With a dc power consumption of 0.5 W, our oscillator achieves FOM values of 199.3 dB, 211.4 dB, and 212.0 dB at 1 kHz, 10 kHz, and 100 kHz offsets, respectively. The power-normalized FOM of our oscillator at a 10-kHz offset is competitive, although it remains below those of highly optimized discrete oscillators (**Table 2**), including temperature-compensated crystal oscillators [66], surface-transverse-wave (STW) oscillators [65], and high-overtone bulk acoustic wave (HBAR) oscillators [67]. The ultralow phase noise of these discrete oscillators generally results from one or more specialized features, including the large $fQ$ product [67] available from HBAR, low insertion loss [65,71], high power handling [71], and thermal compensations implemented either through active feedback circuits [66] or material cuts with near-zero TCF [68,71]. Recent photonic microwave-generation platforms [72-74] demonstrate approximately 20 dB lower phase noise at 10-kHz offset (scaled to 1 GHz). However, their large footprint, high power consumption, and high system complexity necessitate further developments for practical deployment.

Furthermore, continuously tuning the oscillation frequency of our oscillator from 1,000 MHz to 1,007 MHz (**Fig. 4f**) is realized by a combination of voltage-controlled phase shifter and thermal tuning of the oscillator by 18 °C (details in **Methods - Device characterizations**). Our oscillator exhibits a temperature coefficient of frequency (TCF) of -85 ppm/°C, based on a measured frequency shift of 1.535 MHz over this 18°C window.

Several approaches could further improve the phase-noise performance of this architecture. First, the far-offset noise floor could be reduced by improving the IDT transduction efficiency and lowering the ring-resonator insertion loss. For example, the near-critically coupled ring resonator described in **Supplementary Note 4** has $Q_c/Q_{in} = 1.03$ and a lower insertion loss of 19.63 dB at 996.68 MHz. Second, the close-to-carrier phase noise could be reduced through passive temperature compensation. Depositing an additional layer of $SiO_2$, which provides an opposite-sign TCF relative to LN [75], is expected to partially compensate the overall TCF while maintaining the waveguide mode confinement (details in **Supplementary Note 5**). Third, injection locking could further suppress the phase noise at low offset frequencies (< 1 kHz) without affecting the noise performance at larger offset frequencies [76].

For the monolithic integration, the off-chip gain and phase-control elements can be replaced with on-chip counterparts based on acoustoelectric amplification [77] and electro-acoustic control of phononic waveguides [39]. The large footprint of our ring resonator is due to the large ring radius required to maintain low bending loss and high *Q*. Depositing a thicker SiN layer is a promising pathway to reducing the device footprint, as a thicker SiN layer is expected to enhance the acoustic waveguiding contrast, improve mode confinement, and suppress the bending loss. Thicker SiN films, however, require careful stress engineering during deposition, because the increased thermal-expansion mismatch and residual stress between SiN and LN can lead to film cracking.

## Discussion

We demonstrate a low–loss PnICs platform using SiN on LN. Low-insertion-loss SAW building blocks at 1 GHz such as waveguides, directional couplers, and high-*Q* ring resonators are realized (**Table 3**).

Building upon the ring resonator, we further demonstrate a 1-GHz SAW oscillator with a low phase noise of -159.0 dBc/Hz at a 100-kHz offset frequency.

Our slot-waveguide-based circuit architecture is implemented on a bulk piezoelectric substrate, maintaining propagation loss comparable to those of acoustic rib waveguides, while eliminating the needs to develop thin-film technologies [36,78] and specialized etching processes for piezoelectric materials. This approach may facilitate the application of our circuit architecture to emerging material platforms (e.g., potassium niobate.) which often lack sophisticated nanofabrication technologies.

Besides the application of microwave oscillators, our low–loss PnICs platform could also provide opportunities for quantum phononics and hybrid quantum phononic systems. Although our present measurements were performed at room temperature, our prior work had shown that cooling to 5 K can improve the waveguide propagation *fQ* product by approximately 60-fold [39]. Based on this scaling, our waveguide-based acoustic ring resonator could potentially reach a loaded *fQ* product of approximately $1.1\times10^{15}$ Hz at 5 K. This projected value is of the same order of magnitude as the mechanical-mode *fQ* products reported in a diamond optomechanical-crystal resonator ($4.1\times10^{14}$ Hz) [79] and a superconducting-qubit-coupled HBAR ($4.7\times10^{15}$ Hz) [15] , while remains two orders of magnitude below the $2.7\times10^{17}$ Hz achieved in a discrete HBAR optomechanical system [80]. We note that the mechanical *fQ* product alone does not determine the performance of a hybrid quantum system. The more relevant figure of merit is often the cooperativity, which depends jointly on the lifetimes of both interacting modes and their coupling strength. Our waveguide-based PnIC architecture may provide a scalable circuit-level foundation for interfacing acoustic modes with superconducting qubits, optical fields, and solid-state defects.

# Methods

### Device design and fabrication

The device is on an X-cut LN substrate with a SiN thin film deposited on top. The pitch of IDT fingers is 1.651 μm, which corresponds to the half wavelength of acoustic wave at 1.005 GHz. The width of the IDT is 50 μm. The thickness of aluminum is 100 nm. The number of IDT pairs is 35. A 300-nm-thick SiN layer is deposited using plasma-enhanced chemical vapor deposition on the *X*-cut LN substrate. The SiN layer is patterned using electron-beam lithography (EBL) with a 900-nm-thick polymethyl methacrylate (PMMA) resist and etched using reactive-ion etching with carbon tetrafluoride, sulfur hexafluoride and trifluoromethane gases. The metal layer is patterned using EBL with a 300-nm-thick PMMA resist. A 100-nm-thick aluminium layer is deposited using electron-beam evaporation, followed by lift-off in N-Methyl-2-pyrrolidone (NMP).

### Device characterizations

The *S*-parameter spectra measurements are conducted using a Vector Network Analyzer (Keysight P5000A). The positive feedback loop of SAW oscillation consists of our acoustic resonator, a LNA (Mini-circuits, ZKL-33ULN-S+), a phase shifter (RF LAMBDARFPSHT0002W1), a 3-dB attenuator, and a coupler (Mini-circuits, ZFDC10-5-S+). The signal of our oscillator is coupled out from the coupler, and the corresponding phase noise is directly measured using a phase noise analyzer (R&S FSWP).

For the tuning of the oscillation frequency, we use a voltage-controlled phase shifter (Mini-circuits, JSPHS-1000) and a temperature-controlled sample stage. The temperature-controlled sample stage uses a *K*-style temperature sensor and a 6.8-Ω 25-W power resistor (CGS HSA256R8). The temperature sensor is read out by a multimeter (Fluke 45), whereas the power resistor is powered by a d.c. power supply (RIGOL DP832A). A proportional–integral–derivative (PID) control algorithm in Python is used to stabilize the temperature. The temperature stability of the stage is measured under two conditions: free-running operation at room temperature and PID-controlled heating at a setpoint 18 °C above room temperature. Over a 1,000-s measurement interval, the measured peak-to-peak temperature variations were approximately 277 mK and 227 mK, respectively (**Supplementary Fig. 8**). Temperature was monitored at a sampling rate of four samples per second. In this acquisition configuration, the voltage readout resolution was 0.001 mV, corresponding to an equivalent temperature resolution of approximately 25.2 mK. We note that the stage temperature was left uncontrolled under ambient laboratory conditions during the phase-noise measurements. PID control was used only during the oscillation-frequency tuning experiment.

# Acknowledgements

**Funding:** Device fabrication was conducted at the Center for Nanophase Materials Sciences (CNMS2022-B-01473, CNMS2024-B-02643), which is a US Department of Energy Office of Science User Facility. Research was partially supported by the Air Force Office of Scientific Research (AFOSR) under Grant Number W911NF-23-1-0235 and Award Number FA9550-22-1-0548, and by Commonwealth Cybersecurity Initiative in Virginia. Development of the optical vibrometer was partially supported by the Defense Advanced Research Projects Agency (DARPA) OPTIM program under contract HR00112320031. Development of the phononic device was partially supported by DARPA SynQuaNon DO program under Agreement HR00112490314. Measurements of the phase noise were partially supported by DARPA GRYPHON program under Agreement HR00112220008. The views and conclusions contained in this document are those of the authors and do not necessarily reflect the position or the policy of the United States Government. No official endorsement should be inferred. Approved for public release.

**Author contributions:** J.J., J.G.T., and Z.X. contributed equally to this work. J.J. designed the chip with the help of L.S. J.J. fabricated the devices with the help from L.S., Z.X., and Y.Q. J.J. and Z.X. performed the measurements and analyzed the data. Z.X., J.J., R.L., K.F., A.B., X.Y., and L.S. performed phase noise measurements. J.G.T. and Y.Z. performed optical vibrometer measurements. All authors analyzed and interpreted the results. J.J. prepared the manuscript with revisions from all authors. L.S. supervised the project.

**Competing interests:** J.G.T, Y.Z, and L.S. are co-founders and shareholders of Tern Photonics Inc. The other authors declare no competing interests.

**Data availability:** All data needed to evaluate the conclusions in the paper are present in the paper and/or the Supplementary Materials. Additional data related to this paper may be requested from the authors.

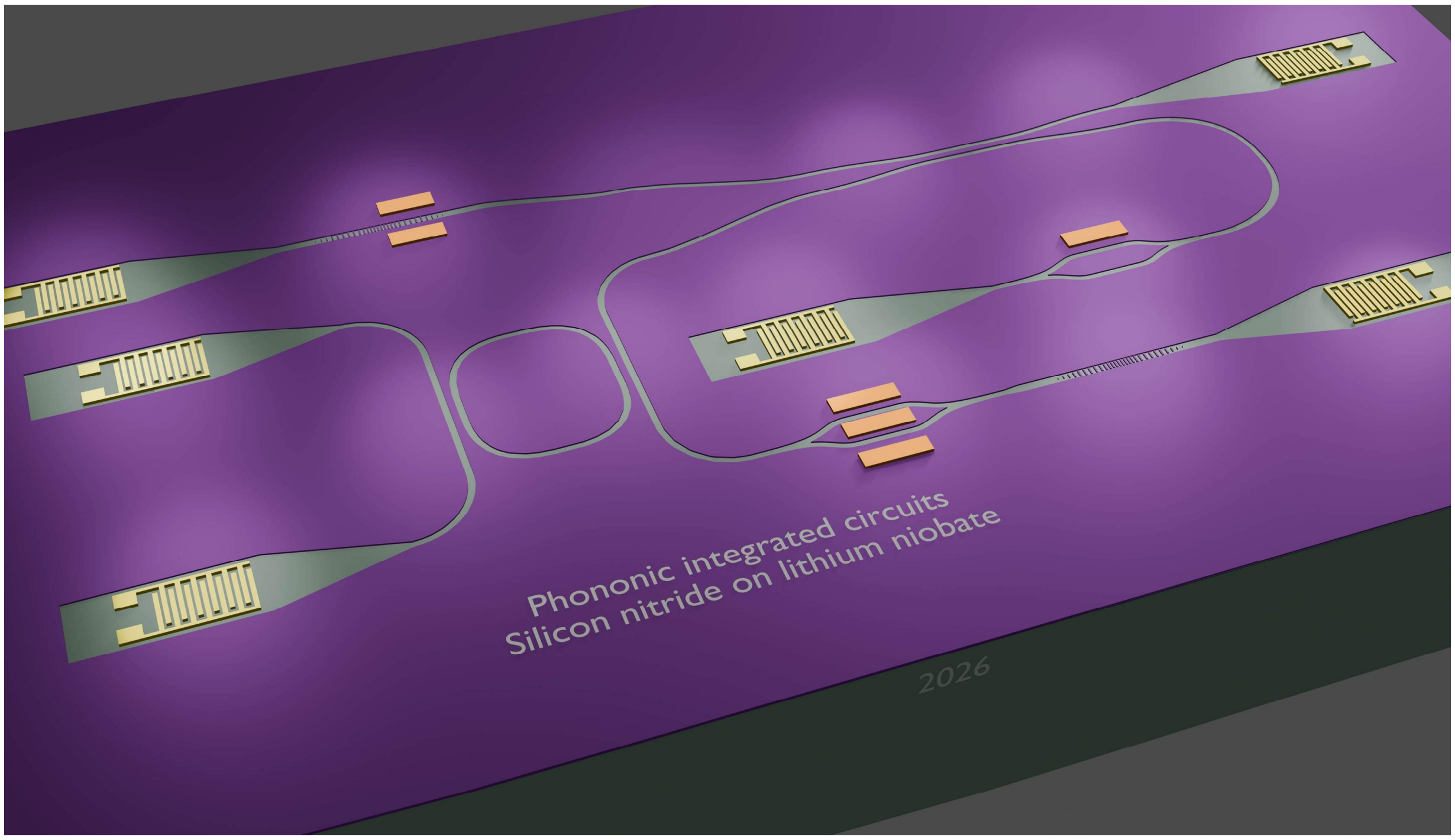

**Fig. 1: Low–loss phononic integrated circuits using SiN on LN.** Schematic of low-loss phononic integrated circuits (PnICs) on a LN substrate (showing green) with a patterned SiN thin film (showing purple) on top. Low-loss SAW building blocks include IDT, waveguides, directional couplers, Mach-Zehnder interferometers, ring resonators, and phononic-crystal resonators. Dimensions are not scaled.

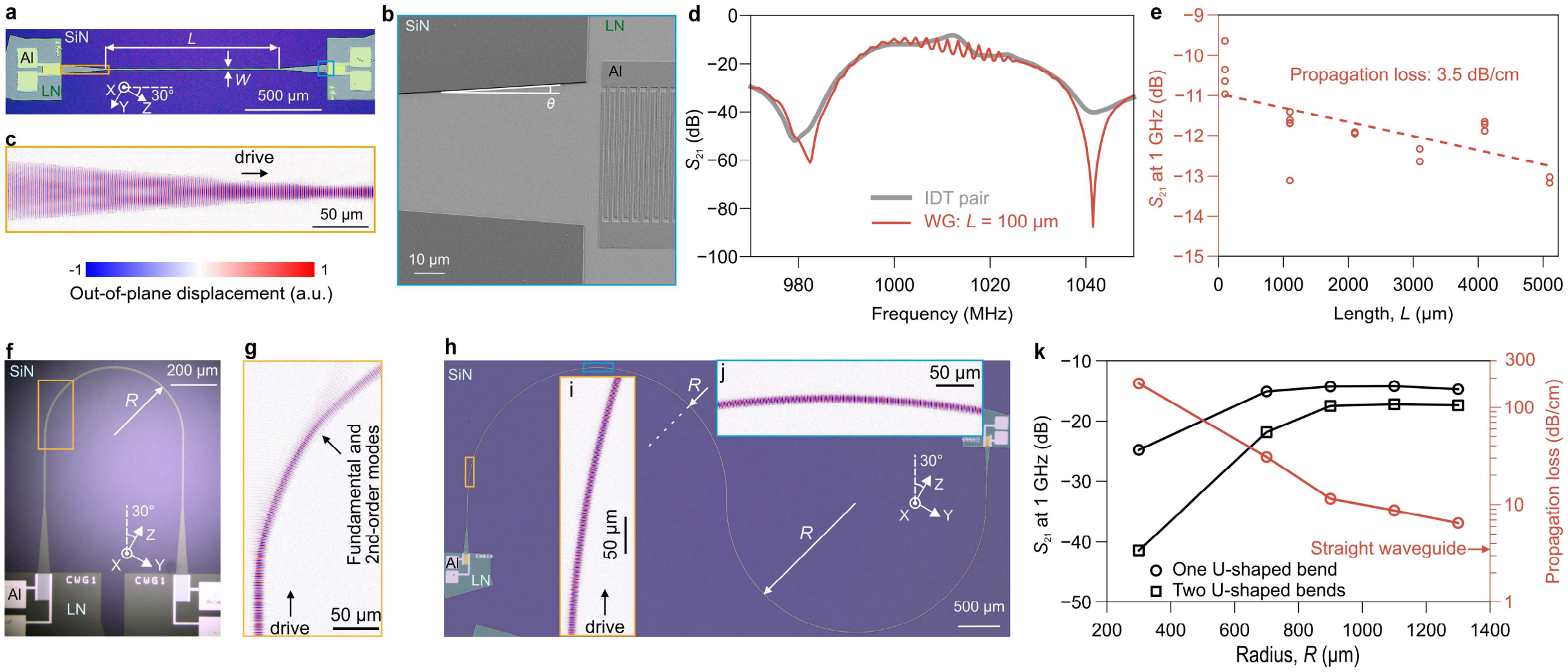

**Fig. 2: Low-loss acoustic waveguides. (a)** Optical micrograph of the fabricated acoustic waveguide with a waveguide width of $W$ and a length of $L$. **(b)** Scanning electron microscopy (SEM) images of the tapering transition between the IDT and the waveguide with a tapering angle of $\theta$. **(c)** Measured out-of-plane displacement profile at the tapering region using our in-house optical vibrometer. **(d)** Measured transmission spectrum $S_{21}$ of our acoustic waveguide (WG) with $W$ =10 μm and $L$ = 100 μm, compared with that of a reference interdigital transducer (IDT) pair. **(e)** Extracted propagation loss of 10-μm-wide acoustic waveguides based on transmission spectra $S_{21}$ as a function of the waveguide length $L$. **(f,h)** Optical micrograph of acoustic waveguides with one U-shape bend and two cascaded U-shape bends, respectively. **(g)** Measured out-of-plane displacement profile at the bending region for a bending radius $R$ = 300 μm, showing strong acoustic wave leakage. **(i,j)** Measured out-of-plane displacement profiles for waveguides with a larger bending radius $R$ =1300 μm, demonstrating suppressed leakage. **(k)** Propagation loss of 10-μm-wide bent acoustic waveguides as a function of bending radii $R$. The color scale is independently normalized for **(c)**, **(g)**, **(i)**, and **(j)**. Al, aluminum; SiN, silicon nitride; LN, lithium niobate; a.u., arbitrary units.

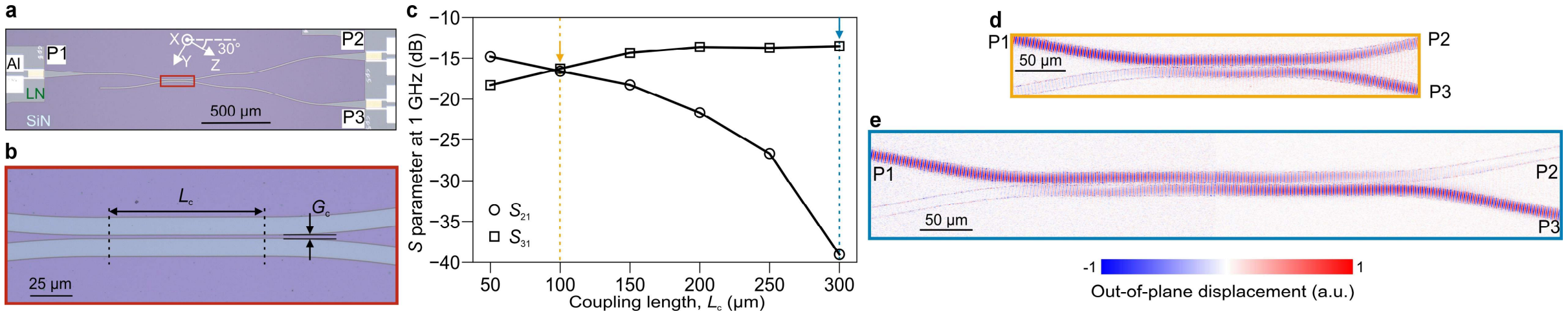

**Fig. 3: Low-loss acoustic directional couplers. (a)** Optical micrograph of the fabricated acoustic directional coupler on lithium niobate. **(b)** Close-up view of the coupling region, showing the coupling gap $G_c$ and coupling length $L_c$. **(c)** Measured transmission $S_{21}$ ($S_{31}$) at 1 GHz for the through (drop) port P2 (P3), as a function of coupling length $L_c$, demonstrating controllable power transfer between the two waveguides. **(d)** Measured out-of-plane displacement profile of a 50/50 directional coupler, where acoustic power is equally split between the through and drop ports. **(e)** Measured out-of-plane displacement profile of a strongly coupled (near-complete power-transfer) directional coupler, where most of the acoustic power is transferred to the drop port. The color scale is independently normalized for **(d–e)**. Al, aluminum; SiN, silicon nitride; LN, lithium niobate; a.u., arbitrary units.

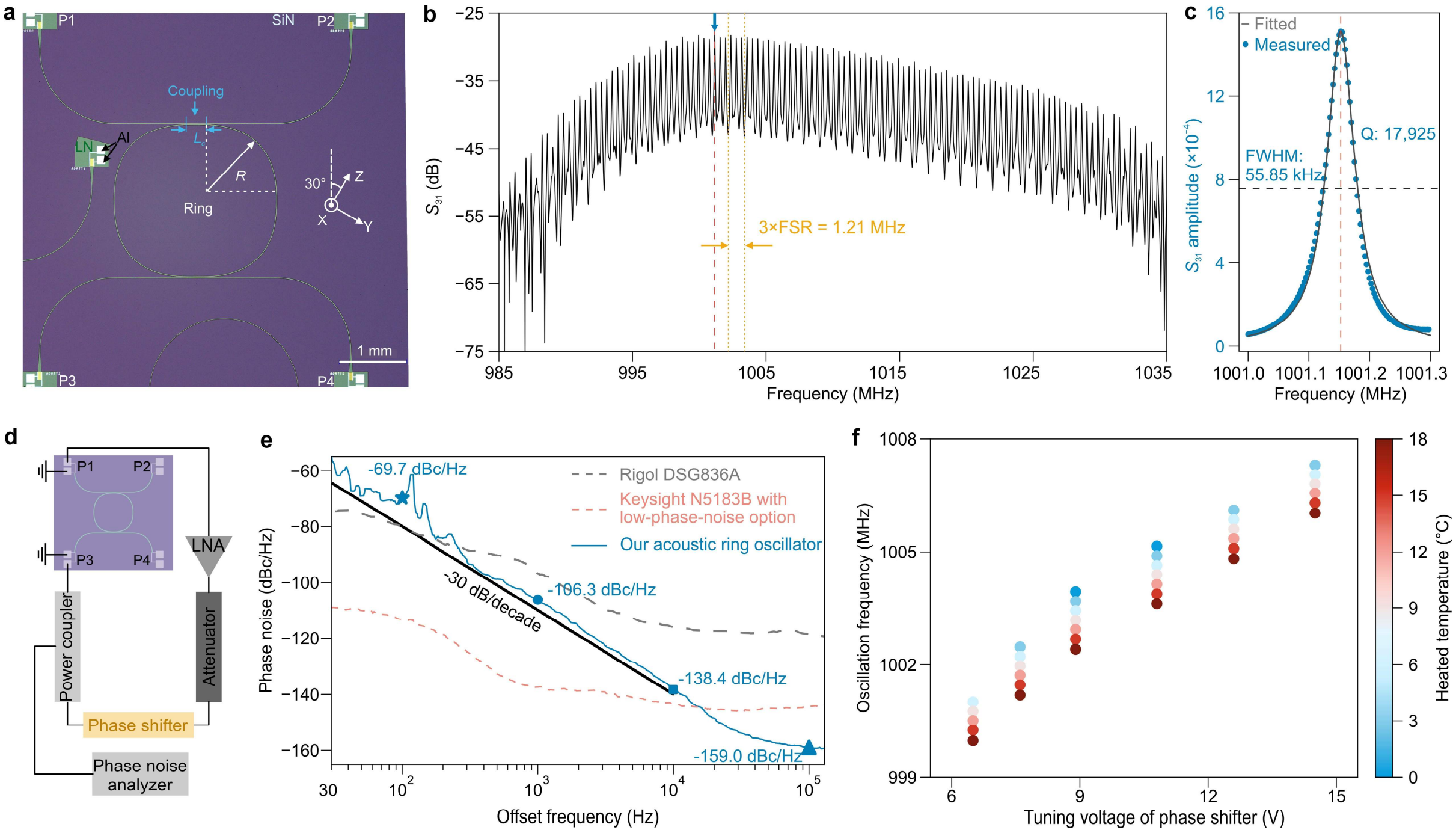

**Fig. 4: High-*Q* acoustic ring resonators and low-phase-noise oscillators. (a)** Optical micrograph of the fabricated four-port ring-like resonator with a radius of $R$ and a coupling length of $L_c$. **(b)** Measured transmission spectrum $S_{31}$ of our ring resonator with $R$ = 1300 μm and $L_c$ = 100 μm. High-*Q* modes are spectrally spaced at a free spectral range (FSR) of 0.403 MHz. **(c)** Close-up view of the measured (dots) and fitted (solid line) $S_{31}$ of the mode with the lowest insertion loss. This mode is centered at 1001.15 MHz with a full width at half maximum (FWHM) of 55.85 kHz, resulting in a loaded quality ($Q$) factor of 17,925. **(d)** Experimental setup used to build and characterize our acoustic ring oscillator. Our oscillator consists of the ring resonator, a low noise amplifier (LNA), an attenuator, a microwave phase shifter and a microwave power coupler. The oscillator outputs from the coupling port of the coupler are measured by a phase noise analyzer. **(e)** Measured phase noises of our acoustic ring oscillator, compared with those from Rigol signal generator DSG836A and Keysight signal generator N5183B with the low-phase-noise option. (**f**) Continuous tuning of the oscillation frequency from 1,000 MHz to 1,007 MHz through a combination of voltage-controlled phase shifter and thermal tuning of the oscillator by 18 °C. Al, aluminum; SiN, silicon nitride; LN, lithium niobate.

**Table 1 Comparisons of key performance metrics for different competing acoustic waveguide and delay line (DL) platforms at room temperature.**

| Platform | | $f_0$ (GHz) | $v_g$ (km/s) | Propagation loss (dB/cm) | Propagation $Q$ | Propagation $f \cdot Q$ | IL* (dB) |
|---|---|---|---|---|---|---|---|
| SiN on LN (this work) | Single mode | 1.0 | 3.4 | 1.9 | 42241 | $4.22 \times 10^{13}$ | ~10 ($L$ = 100 μm) |
| | Multimode | | | 3.5 | 22931 | $2.29 \times 10^{13}$ | |
| GaN-on-sapphire [44] | | 0.2 | 4.18 | 0.5 | 26113 | $5.22 \times 10^{12}$ | >70 |
| GaN-on-SiC [45] | | 3.4 | 3.76 | 36 | 6854.2 | $2.33 \times 10^{13}$ | 38 ($L$ = 4100 μm) |
| LN-on-diamond [36] | | 2.8 | 2.8 | - | - | - | 21 ($L$ = 100 μm) |
| AlScN-on-SiC [37] | | 4.05 | 6.48 | 107 | 1593.9 | $6.46 \times 10^{12}$ | ~ 30 ($L$ = 320 μm) |
| AlScN-on-SiC [42] | | 2.1 | 5.05 | 16.1 | 7048.1 | $1.48 \times 10^{13}$ | 32 ($L$ = 160 μm) |
| AlScN-on-Sapphire DL [55] | | 5.93 | 5.778 | 92 | 3044.1 | $1.81 \times 10^{13}$ | 7.6 ($L$ = 39 μm) |
| Suspended ScAlN DL [81] | | 10.5 | 5.3 | 858 | 630 | $6.62 \times 10^{12}$ | ~30 ($L$ = 50 μm) |
| Suspended LN DL [82] | | 11.5 | 5.29 | 440 | 1348 | $1.55 \times 10^{13}$ | 12.9 ($L$ = 20 μm) |
| LN/a-Si/SiC DL [83] | | ~16.55 | - | - | ~2400 | ~$3.97 \times 10^{13}$ | 28 ($L$ = ~ 50 μm) |
| P3F LN DL [84] | | ~27.8 | - | - | ~1600 | ~$4.40 \times 10^{13}$ | - |

* IL (Insertion loss), here, is defined as $S_{21}$ between two electrical ports of acoustic waveguides or delay lines.

**Table 2 Comparisons of phase noise of different acoustic oscillator designs.**

| Ref. | Materials | Acoustic elements | $f_0$ (MHz) | $f_0Q$† (THz) | IL* (dB) | Phase noise (dBc/Hz) | | | | | | FOM (dB) |
|---|---|---|---|---|---|---|---|---|---|---|---|---|
| | | | | | | 1 kHz offset | | 10 kHz offset | | 100 kHz offset | | 10 kHz offset |
| This work | SiN/LN | Waveguide-based ring resonator | 1,001.2 | 17.9 | 28.2 | -106.3 | -106.3 | -138.4 | -138.4 | -159.0 | -159.0 | 211.4 |
| [60] | LN | SAW resonator | 1026 | 2.87 | 20 | -102.0 | -102.2 | -132.5 | -132.7 | -140.0** | -140.2 | 205.7 |
| [70] | LN/$SiO_2$/Sapphire | SH-SAW delay line | 905 | - | 11 | -104.8 | -103.9 | -130.7 | -129.8 | -149.3 | -148.4 | 210.0 |
| [85] | LNOS | SH-SAW delay line | 888.6 | 0.273 | 5.2 | -94.1 | -93.1 | -117.5 | -116.5 | -139.2 | -138.2 | 196.6 |
| [86] | ZnO/diamond | SAW resonator | 1008 | 0.302 | 9.0 | -90.0** | -90.1 | -118.0** | -118.1 | -137.0** | -137.1 | 191.3 |
| [69] | LN | $SH_0$ delay line | 157 | 0.0209 | 3.2 | -101.0 | -84.9 | -127.0 | -110.9 | - | - | 182.3 |
| [66] | Crystal | - | 26 | - | - | -138.0 | -106.3 | -154.0 | -122.3 | -159.0 | -127.3 | 216.1 |
| [77] | InGaAs/LN# | SAW resonator | 999 | 0.0959 | 25.0** | -57.0 | -57.0 | -82.0** | -82 | -110.0* | -110.0 | 166.7 |
| [65] | - | STW resonator | 1000 | - | 5.8 | -119.0 | -119.0 | -142.0 | -142.0 | -156.0** | -156.0 | 220.9 |
| [71] | Quartz | STW resonator | 1000 | 2.74 | 3.6 | -107.0** | -107.0 | -137.0** | -137.0 | -167.0** | -167.0 | - |
| [68] | Quartz | SAW resonator | 500 | 3.00 | - | -135.0** | -129.0 | -160.0** | -154.0 | -175.0** | -169.0 | - |
| [87] | Silicon | BAW resonator | 103.4 | 8.27 | - | -108.0 | -88.3 | -120.0** | -100.3 | -130.0** | -110.3 | 196.1 |
| [88] | ZnO/sapphire | HBAR | 1200 | - | - | -85.0** | -86.6 | -112.0 | -113.6 | - | - | 204.6 |
| | | | 3670 | | | -80.0** | -91.3 | -100.0 | -111.3 | | | 206.2 |
| [67] | AlN/sapphire | HBAR | 2298 | 55.2 | 17.0 | -115.0 | -122.2 | -143.0 | -150.2 | -153.0** | -160.2 | 217.5 |

The values in the shaded cells are scaled to a 1 GHz carrier.
† Loaded $f_0Q$ is used here. For delay lines, the equivalent loaded $Q$ is calculated from the measured group delay.
* IL (Insertion loss), here, is defined as $S_{21}$ between two electrical ports of acoustic resonators.
**Data extracted from figures.
# This device has solid-state in-cavity gain.

**Table 3 Building blocks of our integrated phononic circuit.**

| Component | Specifications | Insertion loss* at 1 GHz (dB) |
|---|---|---|
| IDT Pair | The IDT finger pitch is 1.651 μm; the IDT width is 50 μm; the thickness of aluminum is 100 nm; the number of IDT pairs is 35. | 10 |
| IDT-Waveguide tapering | $\theta \approx 4.3°$ | <1 |
| Waveguides | $L$ = 0.1 mm, $W$ = 10 μm | ~0.4 |
| Directional coupler | $W$ = 10 μm, $G$ = 2 μm | 13.5-10 = 3.5 between the input and drop ports. |
| Ring resonator | $W$ = 14 μm, $R$ = 1300 μm | 28.2-10 = 18.2 between the input and drop ports; Included 4 dB from the access waveguides. |

* Insertion loss, here, is defined as $S_{21}$ (or $S_{31}$) between two ports of each acoustic component.

Supplementary Materials for

# Low-loss phononic integrated circuits based on a silicon nitride-lithium niobate platform

Jun Ji[†,*], Joseph G. Thomas[†], Zichen Xi[†], Ruxuan Liu, Kinson Fang, Yuan Qin, Andreas Beling, Xu Yi, Yizheng Zhu, and Linbo Shao[*]

[†]J.J., J.G.T., and Z.X. contributed equally to this work.
* Corresponding author. Email: junji@vt.edu; shaolb@vt.edu

**This PDF file includes:**

Supplementary Notes 1-5

Supplementary Figs. 1-8

## Supplementary Note 1. $k^2$ simulation

The electromechanical coupling coefficient is calculated from the phase velocity difference between the cases where the boundary condition at the IDT coupler surface is electrically free or shorted:

$$k^2 = \frac{2(v_{\text{free}} - v_{\text{shorted}})}{v_{\text{free}}} \tag{S1}$$

A large $k^2$ means higher conversion efficiency between electrical and acoustic energy in a piezoelectric material. According to our simulations, $v_{\text{free}}$ = 3388.9 m/s and $v_{\text{shorted}}$ = 3317.2 m/s, leading to a $k^2$ of 4.2%.

## Supplementary Note 2. Temporal coupled-mode theory for acoustic ring resonators

We assume a circulating field amplitude $A$ in the resonator and $S_i$ in the waveguide port (Pi) (**Fig. 4a**). Specifically, $S_1$ is the input port (P1), $S_2$ is the through port (P2), and $S_3$ is the coupled port (P3). We can formulate the standard coupled mode equations as follows:

$$\frac{dA}{dt} = -i\omega_0 A - 2\frac{\gamma_c}{2}A - \frac{\gamma_{\text{in}}}{2}A + \sqrt{\gamma_c}S_1 \tag{S2}$$

$$S_2 = S_1 - \sqrt{\gamma_c}A \tag{S3}$$

$$S_3 = \sqrt{\gamma_c}A \tag{S4}$$

$\omega_0$ is the resonant frequency of our resonator; $\gamma_c = \omega_0/Q_c$ is the loss rate with each coupler; $\gamma_{\text{in}} = \omega_0/Q_{\text{in}}$ is the intrinsic loss rate with our resonator. For harmonic excitation at a frequency $\omega$, we can solve these equations to extract the power transmission $S_{31}$:

$$S_{31}(\omega) = \eta\frac{|S_3|^2}{|S_1|^2} = \eta\frac{\gamma_c^2}{(\omega-\omega_0)^2+\left(2\frac{\gamma_c}{2}+\frac{\gamma_{\text{in}}}{2}\right)^2} \tag{S5}$$

$\eta = \eta_{\text{idt}}\eta_{\text{ex,wg}}$ is the power efficiency of the acoustic elements external to the waveguide-ring coupling. Here, $\eta_{\text{idt}}$ accounts for the efficiency of the IDT pair (10% in our case), $\eta_{\text{ex,wg}}$ accounts for propagation loss in the access waveguides connecting the IDTs to the waveguide–ring couplers. The peak power transmission at $\omega = \omega_0$ is:

$$S_{31}(\omega_0) = \eta\frac{\gamma_c^2}{\left(2\frac{\gamma_c}{2}+\frac{\gamma_{\text{in}}}{2}\right)^2} = \eta\frac{(1/Q_c)^2}{\left(1/Q_c + \frac{1}{2Q_{\text{in}}}\right)^2} \tag{S6}$$

And we have the measured loaded $Q$:

$$\frac{1}{Q} = \frac{2}{Q_c} + \frac{1}{Q_{\text{in}}} \tag{S7}$$

Combining Eq. S6 and Eq. S7, we can retrieve the value of $Q_c$ and $Q_{\text{in}}$.

## Supplementary Note 3. Oscillation mode selection

Our ring resonator supports a set of resonances separated by an FSR of 0.403 MHz. We choose the resonance with the lowest insertion loss near 1001.15 MHz as the oscillation mode and tune the feedback-loop phase so that the Barkhausen phase condition is satisfied at this mode. In practice, we first measure the loop phase with a vector network analyzer (VNA), then adjust the in-loop cable length to bring the total phase close to an integer multiple of $2\pi$, and finally use the RF phase shifter for fine tuning. One RF attenuator in the loop sets the loop gain only slightly above threshold, that is, the loop transmission is measured less than 3 dB using VNA. In this way, oscillations of modes with higher insertion loss are prohibited.

The phase separation between the selected mode and nearby competing modes is quantified. As shown in **Supplementary Fig. 5**, among the resonances whose transmission amplitudes are within 3 dB of the selected mode, the smallest phase difference from the selected mode is 5.6°. This phase difference is sufficiently large to be precisely controlled using our RF phase shifter.

## Supplementary Note 4. Near-critically coupled acoustic ring resonators

To show that our ring resonator is not limited to the under-coupling regime, we experimentally demonstrate a near-critically coupled ring resonator with $Q_c/Q_{in}$ = 1.03. For the 996.968-MHz mode, the measured insertion loss and loaded $Q$ are 19.63 dB and 8,097, respectively (**Supplementary Fig. 6**). The retrieved intrinsic $Q_{in}$ is 23,863, while the retrieved coupling $Q_c$ is 24,511 per coupler. The strong coupling is enabled by the directional coupler (Fig. 3) at a 180-μm-long coupling length. Further increasing the coupling length would move the resonator toward over-coupling. The dimensions of the ring are below: the waveguide width $W$ = 10 μm; ring radius $R$ = 1500 μm; the coupling gap $G_c$ = 2 μm; and the coupling length $L_c$ = 180 μm.

## Supplementary Note 5. Passive thermal compensation

To reduce TCF, we propose passive thermal compensation based on a multiple-layer structure ($SiO_2$/SiN/LN), as shown in **Supplementary Fig. 7a**. In the proposed design, a 100-nm-thick silicon dioxide ($SiO_2$) layer is deposited on top of our current SiN-on-LN waveguide. Because $SiO_2$ can provide a positive-valued TCF, it can partially offset the negative TCF of LN. COMSOL simulations (**Supplementary Fig. 7b**) show that the waveguide mode in the proposed structure remains confined. We will experimentally evaluate the total TCF, insertion loss, and $Q$ of ring resonators with the passive thermal compensation in future work.

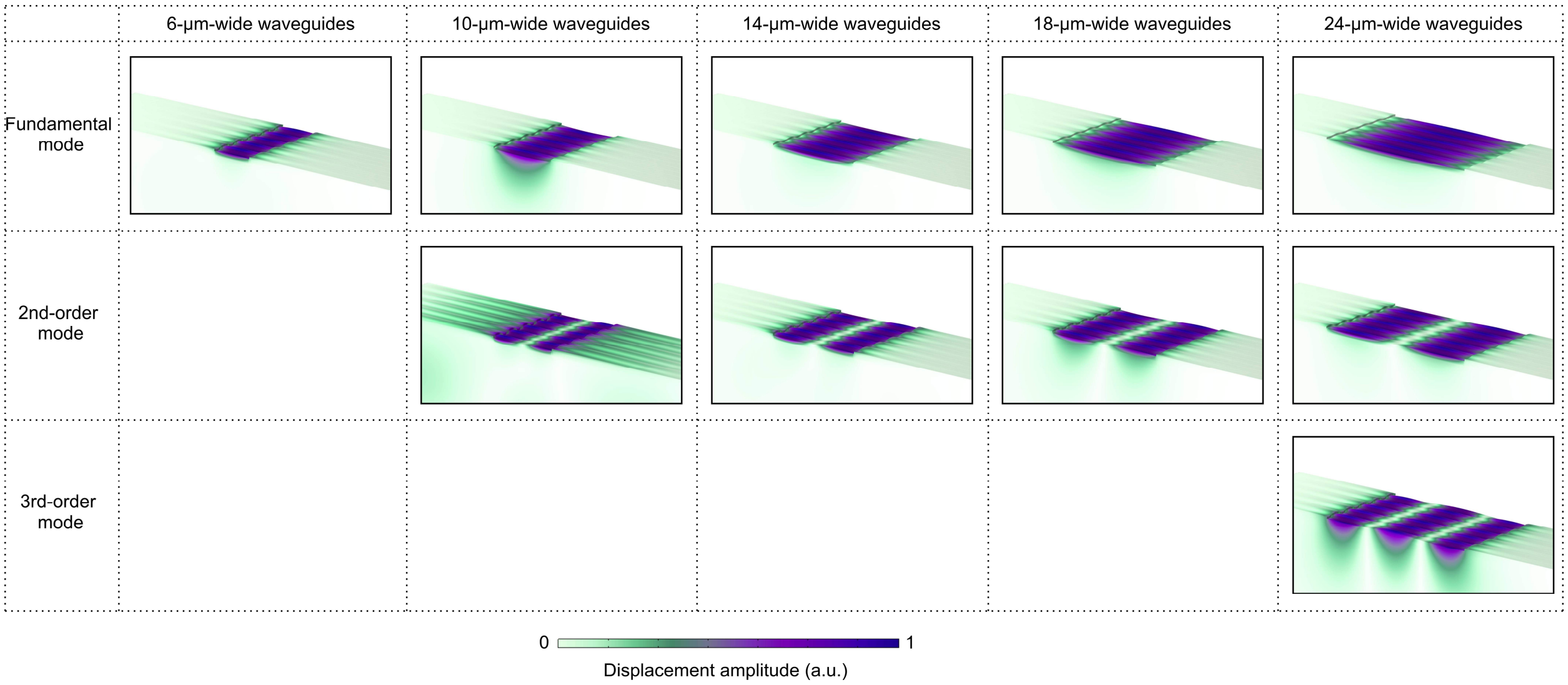

**Supplementary Fig. 1**

**Simulated mode profiles (the total displacement) in our acoustic waveguides with varying waveguide widths *W*.**

For small waveguide width such as $W = 6$ μm, only fundamental mode is supported in the waveguide. For larger waveguide widths such as $W = 10$, 14, 18, and 24 μm, high-order acoustic modes are supported. Acoustic modes are along a 30° angle with respect to the crystal *Z* axis on a *X*-cut LN substrate. The geometry and the color bar of the total displacement amplitude are scaled independently in each subplot.

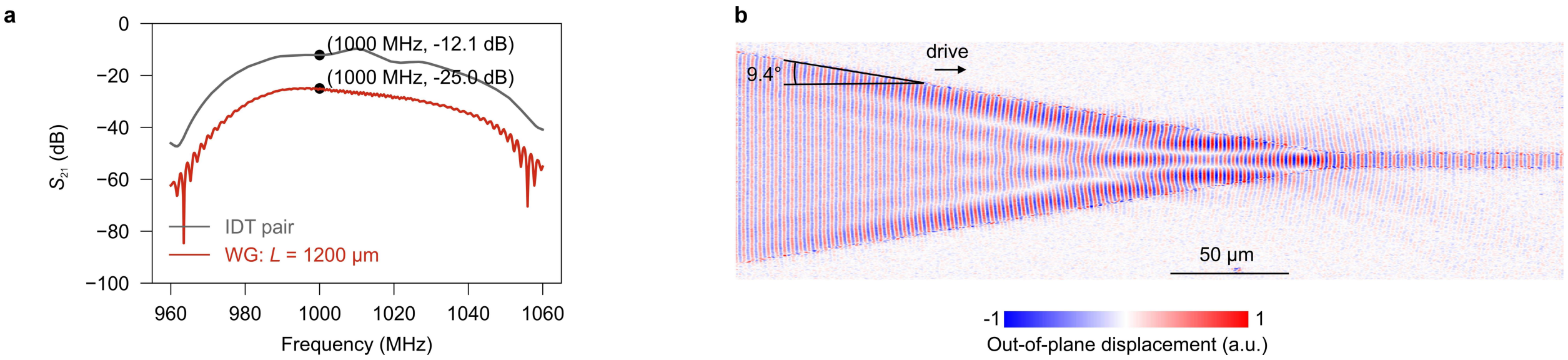

**Supplementary Fig. 2**

**Tapering loss at a large tapering angle.**

**(a)** Measured transmission spectrum $S_{21}$ of an acoustic waveguide with a large tapering angle $\theta$ of 9.4°, compared with that of a reference IDT pair. A large tapering loss of around 12.9 dB is induced by the tapering angle. **(b)** Measured out-of-plane displacement profiles at the tapering region using our in-house optical vibrometer for $\theta$ = 9.4°. Significant acoustic wave scattering is observed at the tapering region, compared with that in **Supplementary Fig. 3**.

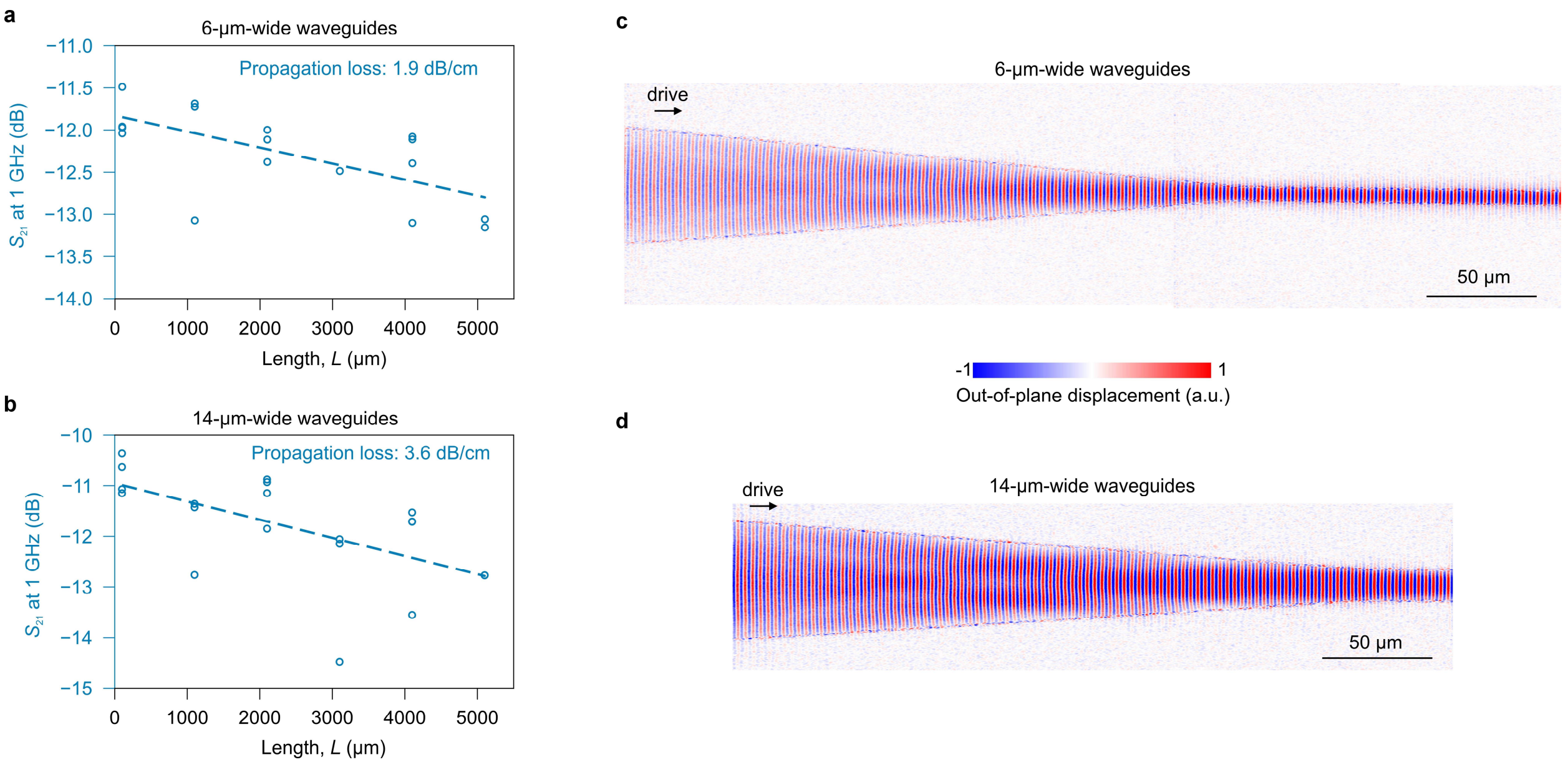

**Supplementary Fig. 3**

**Measured propagation loss and displacement profiles for acoustic waveguides with $W = 6$ and 14 µm.** Measured transmission spectra $S_{21}$ of acoustic waveguides as a function of waveguide length $L$ for a waveguide width of (**a**) $W = 6$ µm and (**b**) $W = 14$ µm and their extracted propagation loss. Measured out-of-plane displacement profiles using our in-house optical vibrometer for acoustic waveguides with a width of (**c**) $W = 6$ µm and (**d**) $W = 14$ µm. The color scale is independently normalized for **(c)** and **(d)**.

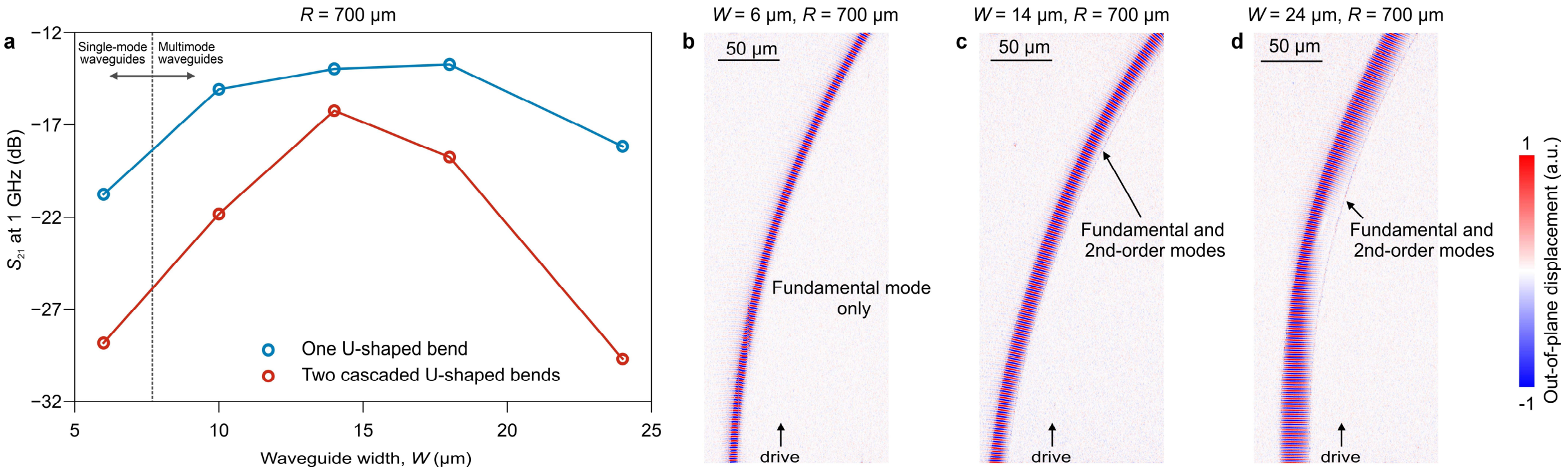

**Supplementary Fig. 4**

**Measured propagation loss and displacement profiles for bent acoustic waveguides (*R* = 700 μm) with different waveguide widths.**

**(a)** Measured transmission spectra $S_{21}$ of acoustic waveguides as a function of waveguide width *W* when the waveguide radius *R* = 700 μm. Measured out-of-plane displacement profiles using our in-house optical vibrometer for acoustic waveguides with a width of (**b**) *W* = 6 μm, (**c**) *W* = 14 μm, and (**d**) *W* = 24 μm. The color scale is independently normalized for **(b-d)**.

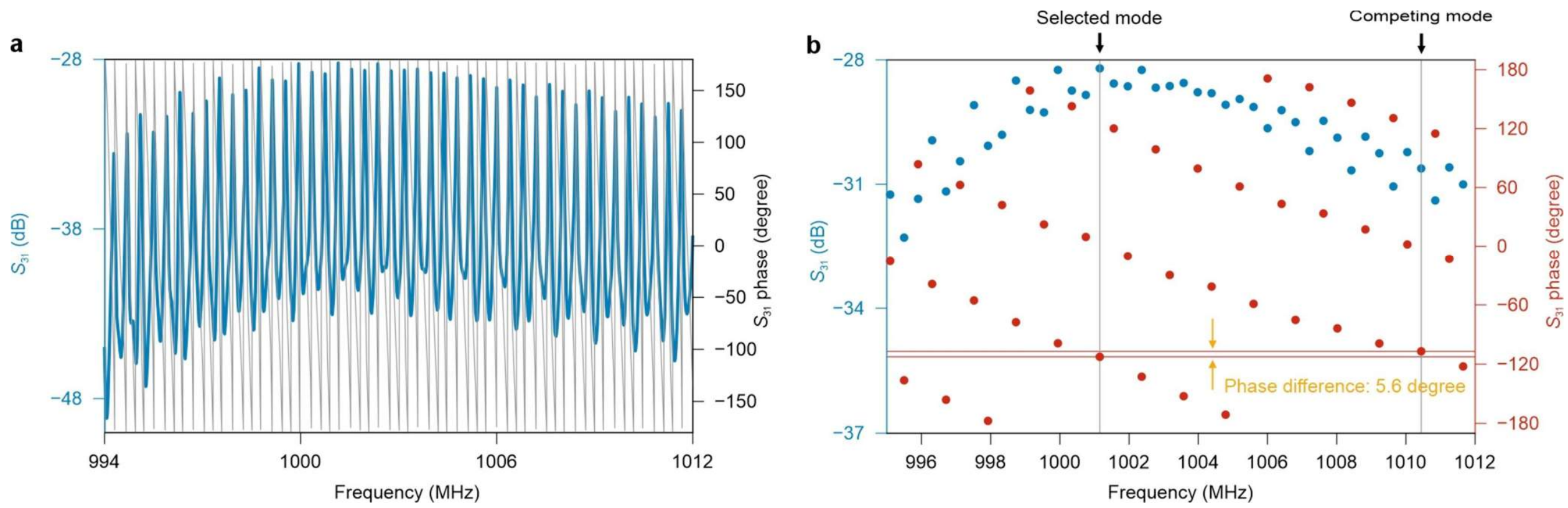

**Supplementary Fig. 5**

**Oscillation mode selection. (a)** Measured transmission spectrum $S_{31}$ and phase response of our ring resonator. **(b)** Measured transmission amplitude and phase of resonance modes near the selected mode. Among all modes whose transmission amplitude is within 3 dB of the selected mode, the smallest measured phase separation from the selected mode is 5.6°. This phase separation is sufficiently large to be resolved and controlled using the RF phase shifter, avoiding mode hopping.

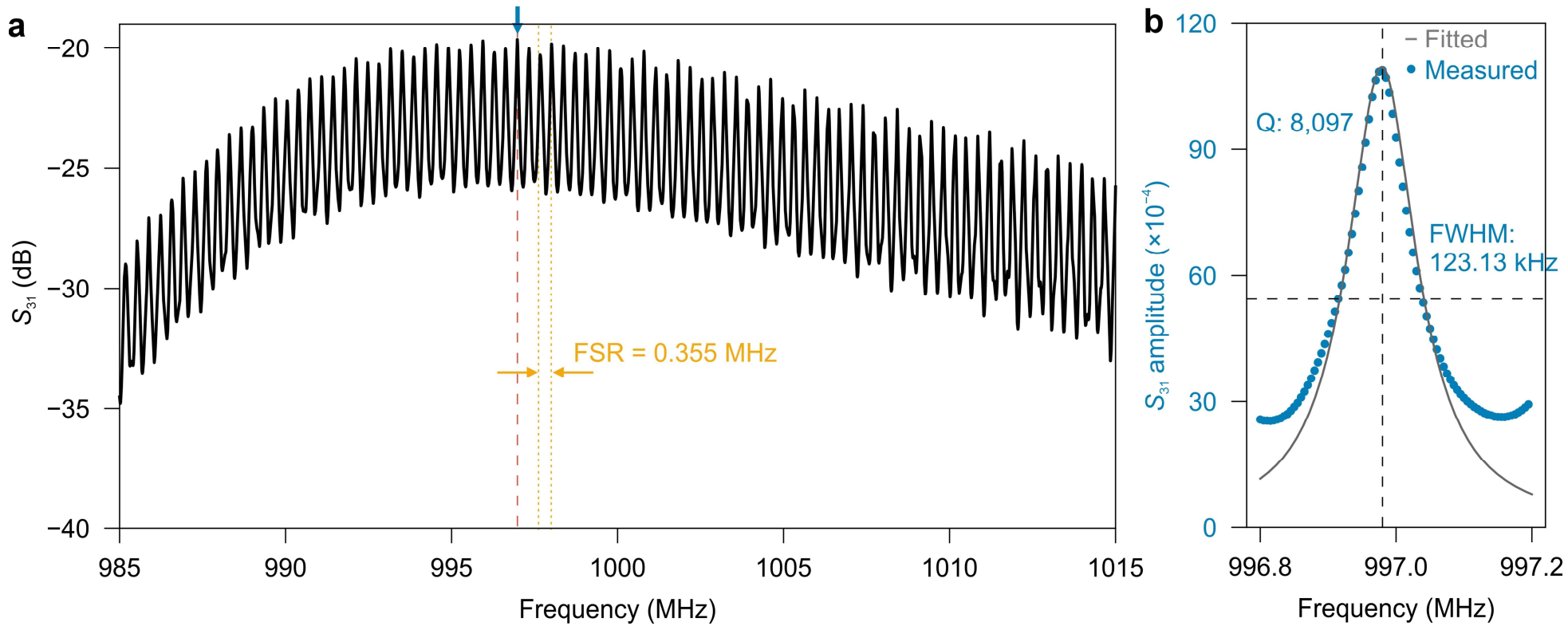

**Supplementary Fig. 6**

**A near-critically coupled ring resonator with $Q_c/Q_{in}$ = 1.03.**

**(a)** Measured transmission spectrum $S_{31}$ of our ring resonator with $R$ = 1500 μm, the coupling gap $G_c$ = 2 μm, and the coupling length $L_c$ = 180 μm. High-$Q$ modes are spectrally spaced at a free spectral range (FSR) of 0.355 MHz. **(b)** Close-up view of the measured (dots) and fitted (solid line) $S_{31}$ of the mode with the lowest insertion loss. This mode is centered at 996.968 MHz with a full width at half maximum (FWHM) of 123.13 kHz, resulting in a loaded quality ($Q$) factor of 8,097. Waveguide width is 10 μm.

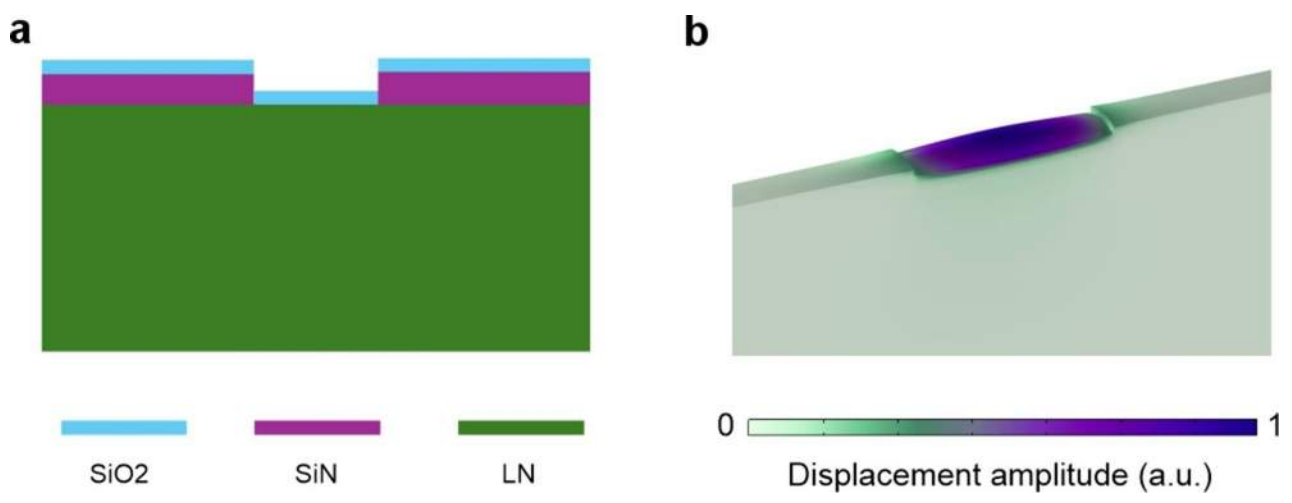

**Supplementary Fig. 7**

**Passive thermal compensation based on a multiple-layer structure ($SiO_2$/SiN/LN).**

**(a)** An additional 100-nm-thick silicon dioxide ($SiO_2$) layer is proposed to be deposited on top of our SiN-on-LN waveguide. **(b)** Simulated displacement profile of the acoustic waveguide mode in the proposed multiple-layer structure. The waveguide mode remains confined.

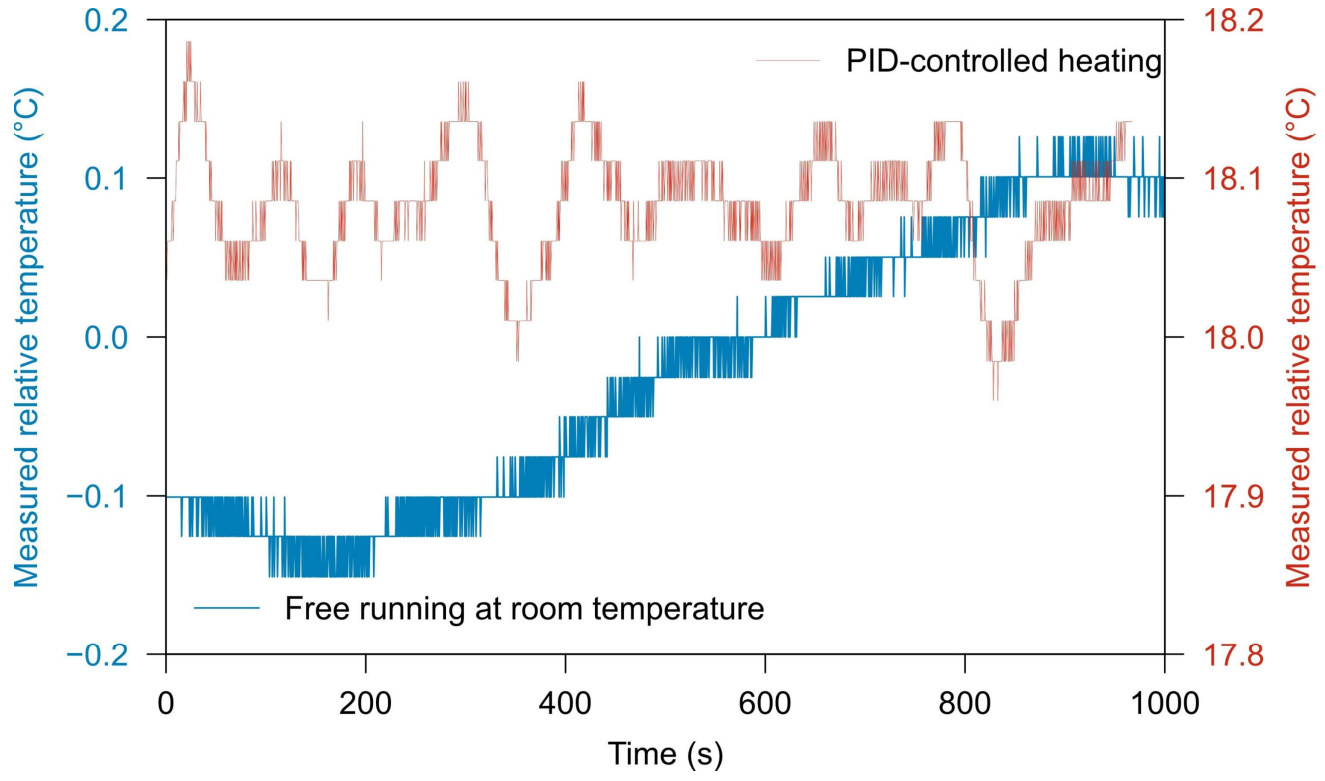

**Supplementary Fig. 8**

**Temperature stability of the stage under free-running and PID-controlled conditions.**

Measured stage temperature relative to the room temperature over 1,000 seconds during free-running operation at room temperature (blue, left axis) and PID-controlled heating at a setpoint 18 °C above room temperature (red, right axis). The corresponding peak-to-peak temperature variations are approximately 277 mK and 227 mK, respectively. Measurements were acquired at four samples per second using a type-K thermocouple and a Fluke 45 multimeter, with an equivalent temperature readout resolution of approximately 25.2 mK.